\documentclass{ws-p10x7}
\usepackage{citesort}
\usepackage{mcite}

\begin{document}

\title{Diffractive Vector Meson Production at HERA}

\author{Bruce Mellado }

\address{{\rm On behalf of the H1 and ZEUS Collaborations}.\\Columbia University, 538 120th Street West, New York, NY 10027, USA\\E-mail: mellado@nevis1.nevis.columbia.edu}

\twocolumn[\maketitle\abstract{The H1 and ZEUS collaborations at HERA report new data on the diffractive production of vector mesons with a many-fold increase of luminosity compared to previous measurements. The new data include photoproduction and electroproduction of $\rho^0,\phi,J/\psi$. The available data on elastic VM production indicate that the interaction scales as $Q^2+M_{V}^2$.}]

\section{Introduction}
\vspace*{-0.3cm}
The production of vector mesons (VM, $V$) at HERA is interesting for the study of non-perturbative hadronic physics, perturbative QCD (pQCD) and their interplay. Moreover, VM production is complementary to deep-inelastic scattering (DIS). DIS has shown the correctness of pQCD down to low values of $Q^2\sim$ 1 GeV$^2$ (where $Q^2$ is the virtuality of the photon exchanged), a region where a transition to non-perturbative physics is observed~\cite{f2feno}. In the past the elastic processes were mostly treated through non-perturbative methods. These methods are successful to describe basic features of exclusive light VM production at low photon virtualities. However, in recent years a pQCD picture of the exclusive VM production is able to describe its basic features provided that $Q^2/\Lambda^2_{QCD}\gg$ 1 and $M_{V}/W\ll$ 1, where $M_{V}$ is the mass of the VM and  $W$ is the photon-proton center-of-mass energy. The photon fluctuates into a quark-antiquark pair, long before the interaction with the proton occurs. The interation occurs via gluon ladders. The steepness of the rise of the VM production with $W$ is basically driven by the gluon density in the proton, which is probed at an effective scale $K\sim(Q^2+M^2_{V})$ at low $x$ ($x\simeq(Q^2+M_V^2)/W^2$).

Assuming a flavor independent production mechanism the relative production rates should scale approximately with the square of the quark charges, i.e. the relative production rates scale as $\rho^0:\omega:\phi:J/\psi=9:1:2:8$, referred here as SU(4) ratios. It is interesting to determine how the interaction changes the SU(4) ratios.

As for today the H1 and ZEUS collaborations have measured the elastic production of vector mesons $e+p\rightarrow e+V+p$, where $V=\rho^0,\omega,\phi,J/\psi$ over a wide range of $W$, from photoproduction ($Q^2\simeq 0$) to $Q^2=100$ GeV$^2$ ~\cite{rhophoto,omegaphoto,phiphoto,jpsiphoto,psiphoto,rhoelec,omegaelec,phielec,jpsielec}. New results on proton-dissociative diffractive photoproduction of VM at $W\simeq 100$ GeV and $-t\leq$ 12 GeV$^2$~\cite{hight}, where $t$ is the squared four-momentum transfer at the proton vertex, are also available. The kinematic range of DIS and VM production now overlap at HERA giving us the chance to examine more deeply fundamental issues of the physics of hadronic interactions.

\vspace*{-0.45cm}
\section{New results}
\vspace*{-0.2cm}
The ZEUS collaboration has presented~\cite{rhoelec} new data on the elastic electroproduction of $\rho^0$ using an integrated luminosity of 38 ${\rm pb}^{-1}$. The cross section for $\gamma^\ast p\rightarrow\rho^0 p$ has been measured for $32<W<160$ GeV and $5<Q^2<80$ GeV$^2$. The $W$ dependence of $\sigma(\gamma^\ast p\rightarrow\rho^0 p)$ is measured. If one assumes the form $W^\delta$, where $\delta$ is extracted from a fit at fixed $Q^2$, it is shown that $\delta$ displays a marked increase with $Q^2$. The $Q^2$ dependence of $\gamma^\ast p\rightarrow\rho^0 p$ is fit to a form $(Q^2 + m^2_\rho )^{-n}$. The power $n$ is found to depend on $Q^2$.

\begin{figure}
\epsfxsize120pt
\vspace*{-0.6cm}
\figurebox{120pt}{100pt}{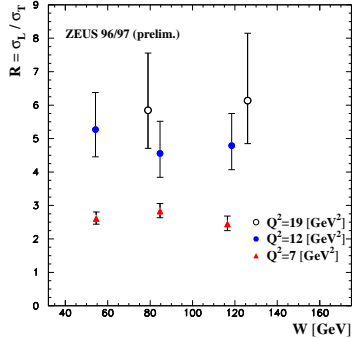}
\vspace*{-0.5cm}
\caption{$R=\sigma_L/\sigma_T$ for $\rho^0$ production is presented for fixed $Q^2$ values. No marked $W$ dependence is observed.}
\label{fig:radk}
\end{figure}

The angular distribution of the decay products of the $\rho^0$ are used to measure the ratio of the production cross sections, $R=\sigma_L/\sigma_T$, where $\sigma_L$ and $\sigma_T$ correspond to longitudinally and transversely polarized virtual photons. The $Q^2$ dependence of $R$ is parametrized as $R={1\over \xi}(Q^2/m^2_\rho)^\kappa$, where $\xi=2.17\pm0.07$ and $0.75\pm0.03$. The $W$ dependence of $R$ is measured for three fixed values of $Q^2$. Fig.~\ref{fig:radk} shows that $R$ does not depend on $W$, therefore the steepness of the rise of $\sigma(\gamma^\ast p\rightarrow\rho^0 p)$ with $W$ appears to be independent of the photon polarization.


The ZEUS collaboration has presented~\cite{jpsiphoto} new results on the elastic photoproduction of $J/\psi$ meson using two decay channels with an integrated luminosity of 38 ${\rm pb}^{-1}$ and 48 ${\rm pb}^{-1}$. The total $\sigma(\gamma p\rightarrow J/\psi p)$ has been measured in the kinematic range $20<W< 290$ GeV, exhibiting a steep dependence on $W$, which is understood within pQCD as being driven by the rise of the gluon density in the proton at low $x$. The differential cross-section $d\sigma /dt$ has been measured in the kinematic range $30<W<170$ GeV and $-t \leq$ 1.8 GeV$^2$. The $d\sigma /dt$ data is used to determine $\alpha_{I\!\!P}(t)$ in six bins of $t$ by means of a fit of the form $d\sigma /dt\propto (W^2)^{2\alpha_{I\!\!P}(t)-2}$. The resulting values of $\alpha_{I\!\!P}(t)$ are fitted to a line, $\alpha_{I\!\!P}(t)=\alpha_{I\!\!P}(0)+\alpha_{I\!\!P}^{\prime}\cdot t$, where $\alpha_{I\!\!P}(0)=1.193^{+0.019}_{-0.015}$ and $\alpha_{I\!\!P}^{\prime}=0.105^{+0.033}_{-0.031}$. The measurement of the intercept, $\alpha_{I\!\!P}(0)$, is neither compatible with the {\it soft} Pomeron nor the {\it hard} Pomeron alone. The slope, $\alpha_{I\!\!P}^{\prime}$, is significantly smaller than that of the {\it soft} Pomeron and suggests the presence of small shrinkage.

\begin{figure}
\epsfxsize150pt
\vspace*{-1.2cm}
\figurebox{120pt}{160pt}{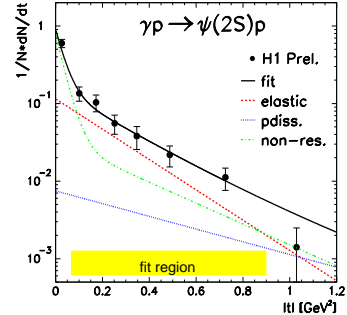}
\vspace*{-1.5cm}
\caption{The corrected number of events $1/N\cdot dN/dt$ of $\psi(1S)$ candidates as a function of $\mid t\mid$. The lines correspond to different contributions to the combined fit of the $t$-slope.}
\label{fig:psi}
\end{figure}

The H1 collaboration has presented~\cite{psiphoto} new results on the elastic photoproduction of $\psi(2S)$ corresponding to an integrated luminosity of 38 ${\rm pb}^{-1}$ in a range $40<W<150$ GeV. The $t$-dependence of the elastic production of $\psi(2S)$ differential cross section is extracted by means of a combined fit that takes into account the contribution of the proton dissociative and non-resonant QED background (see Fig.~\ref{fig:psi}). A single exponential $e^{-b\mid t\mid}$ is used to describe the elastic cross section. The fit yields $b_{\psi(2S)}=(4.5^{+1.7}_{-1.4})$ GeV$^{-2}$. This result is consistent with the slope parameter of the $J/\psi$ of $b_{J/\psi}=(4.73^{+0.39}_{-0.46})$ GeV$^{-2}$~\cite{jpsiphoto}. This is consistent with the pQCD prediction despite the naive expectation that $\psi(2S)$ has more peripheral $t$-dependence since the two $c$-quarks are on average further apart from each other than in the ground state.

The ZEUS collaboration has presented~\cite{jpsielec} new results on the elastic electroproduction of $J/\psi$ using an integrated luminosity of 75 ${\rm pb}^{-1}$ in the kinematic range $50<W<150$ GeV and $2<Q^2<100$ GeV$^2$. The $W$ dependence of the cross section is parametrized with $W^\delta$ at fixed $Q^2$. The obtained values of $\delta$ are consistent with those measured in photoproduction. The $Q^2$ dependence of the cross section is fairly well described by predictions of pQCD. The ratio of the $J/\psi$ to $\rho^0$ cross sections rises rapidly with $Q^2$ approaching the SU(4) ratios. The ratio $R=\sigma_L/\sigma_T$ is extracted from the angular distributions of the decay products of the $J/\psi$ and increases with $Q^2$. The measured $R$ is consistent with the one obtained for $\rho^0$ after taking into account the suppression factor of $M_{\rho}/M_{J/\psi}$.

\begin{figure}
\epsfxsize165pt
\figurebox{120pt}{160pt}{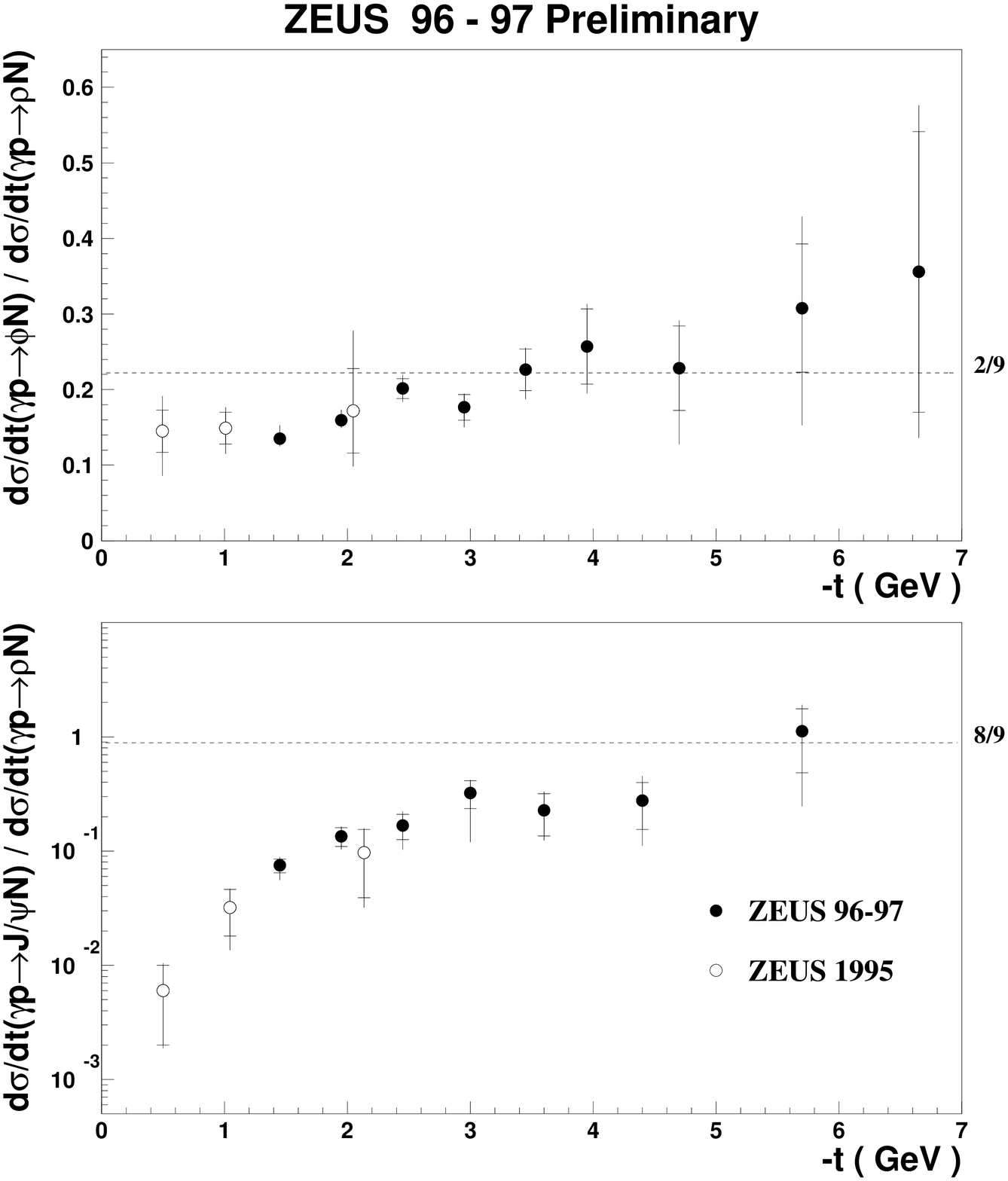}
\caption{The ratios of the cross sections $d\sigma /dt$ for $\phi$ to $\rho^0$ and $J/\psi$ to $\rho^0$ for proton-dissociative photoproduction.}
\label{fig:ht}
\end{figure}

The ZEUS collaboration has produced~\cite{hight} new results on the photoduction of $\rho^0,\phi,J/\psi$ where the proton dissociates into a low mass state $N$. The data correspond to 24 ${\rm pb}^{-1}$ at $W\simeq 100$ GeV and extended up to $-t=12$ GeV$^{-2}$. The measured differential cross section $d\sigma /dt$ is parametrized as a power function $d\sigma /dt\propto (-t)^{-n}$. The fitted value of $n$ decreases with increasing mass of the VM: $n_{\rho}=3.31^{+0.12}_{-0.12}, n_{\phi}=2.77^{+0.18}_{-0.18}$ and  $n_{J/\psi}=1.7^{+0.28}_{-0.28}$. A comparison of the measured differential cross sections with QCD models shows that the perturbative part of the calculation for $\rho^0$ and $\phi$ production at the $t$ values covered in this analysis is well below the data. The pQCD prediction is in agreement with the $J/\psi$ data, but large theoretical uncertainties remain. As illustrated in Fig.~\ref{fig:ht}, the differential cross section ratio ${d\sigma \over dt}(\gamma p\rightarrow\phi N)/{d\sigma \over dt}(\gamma p\rightarrow\rho^0 N)$ increases with -t and approaches the SU(4) ratio at $-t\approx 3.5$ GeV$^2$, while in elastic electroproduction regime the $\phi /\rho^0$ ratio approaches the SU(4) value at a larger scale, $Q^2\geq 6$ GeV$^2$. In the case of the $J/\psi$ this ratio is significantly smaller than the SU(4) value up to $-t$ values of 4 GeV$^2$. For -t$\approx 3.5$ GeV$^2$ this ratio is a factor of five above the $(J/\psi)/\rho^0$ ratio for $Q^2\approx 3.5$ GeV$^2$, indicating that $t$ and $Q^2$ are not equivalent scales.

The decay angle analysis is performed for the $\rho^0$ and $\phi$ and demonstrates a clear deviation from $S$ channel helicity conservation (SCHC). The measured values of the spin density matrix are in agreement with pQCD predictions.

\vspace*{-0.4cm}
\section{The scale of interaction in elastic VM production.}
\vspace*{-0.2cm}
Recent results on the elastic VM production at HERA have shown that the $W$ dependence of the total cross section $\sigma(\gamma^{\ast}p\rightarrow Vp)$ and the $t$ dependence of the differential cross section $d\sigma(\gamma^{\ast}p\rightarrow Vp)/dt$ is dependent on $Q^2$ and $M^2_V$, suggesting that the observables of the interaction be functions of these two variables~\cite{talk} (See other relevant publications~\cite{clerbaux,phielec,naroska}). 

\begin{figure}
\epsfxsize189pt
\vspace*{-1.2cm}
\figurebox{120pt}{160pt}{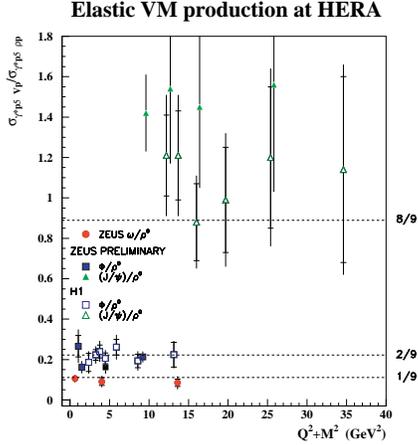}
\vspace*{-1.7cm}
\caption{The total cross section ratios $\sigma_{\omega},\sigma_{\phi},\sigma_{J/\psi}$ to $\sigma_{\rho^0}$ as a function of $Q^2+M_V^2$ at fixed $W$ compared to the SU(4) ratios.} 
\label{fig:rat}
\end{figure}

It has been shown that the relative production ratios reach approximately the SU(4) ratios at $Q^2\gg M_V^2$. However, Fig.~\ref{fig:rat} shows the total cross section ratios $\sigma_{\omega},\sigma_{\phi},\sigma_{J/\psi}$ to $\sigma_{\rho}$ being approximately constant with $Q^2+M_V^2$. The ratio $\sigma_{J/\psi}/\sigma_{\rho}$ in Fig.~\ref{fig:rat} is systematically higher than the SU(4) ratio, possibly due to difference in the wave function of light and heavy quarks.

\begin{figure}
\epsfxsize189pt
\figurebox{120pt}{160pt}{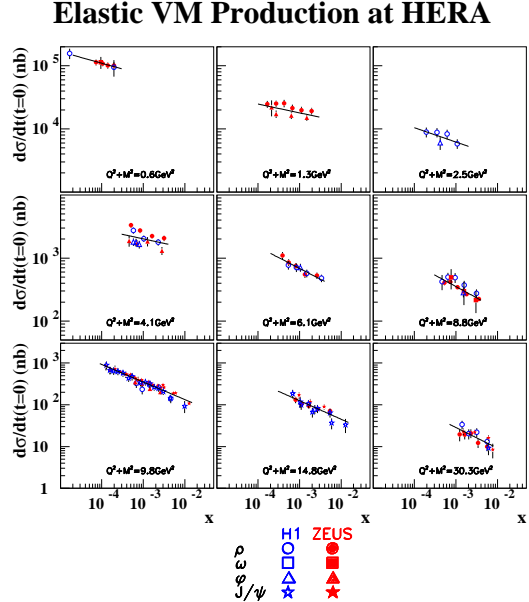}
\vspace*{-1.2cm}
\caption{Elastic $d\sigma /dt(t=0)$ for various VM as a function of $x$ in bins of $Q^2+M^2_V$ (See text). The solid line are fits of the form $A\cdot x^{-2\cdot\lambda_{V}}$.}
\label{fig:xdep}
\end{figure}

The universal character of the interaction is further demonstrated in Fig.~\ref{fig:xdep}, where $d\sigma(\gamma^{\ast}p\rightarrow Vp)/dt(t=0)\approx b_{V}\cdot \sigma(\gamma^{\ast}p\rightarrow Vp) \cdot r^{SU(4)}_{\rho}/r^{SU(4)}_{V}$ is plotted as a function of $x$. Here $b_{V}$ is extracted from the differential cross section $d\sigma(\gamma^{\ast}p\rightarrow Vp) /dt$ assuming $d\sigma /dt\propto e^{-b\mid t\mid}$ and $r^{SU(4)}_{V}$ are the SU(4) ratios. This indicates that the effects due to the wave functions of the different VM do not play a major role. The $x$ dependence of the VM cross sections is parametrized as $A\cdot x^{-2\cdot\lambda_{V}}$, where $A$ and $\lambda_{V}$ are fitted in each bin of $Q^2+M^2_V$. The behavior of the steepness of the rise of the VM production cross section as $x\rightarrow 0$ with changing $Q^2+M^2_V$ is similar to the $x$ dependence of the inclusive $F_2^p$ with changing $Q^2$~\cite{f2feno}. These results indicate that the combination $Q^2+M_V^2$ is a good choice of scale of the interaction in elastic VM production.

\vspace*{-0.5cm}
\begin{thebibliography}{99}
\vspace*{-0.1cm}
\bibitem{f2feno} H1 Collab., C. Adloff et al., Nucl. Phys. B497 (1997) 3; ZEUS Collab., J.~Breitweg et al. Phys.Lett. {\bf B443} (1998) 394.
\bibitem{rhophoto}  H1 Collab., S.~Aid et al., Nucl.Phys. {\bf B463} (1996) 3; ZEUS Collab., J.~Breitweg et al., Eur.Phy.J. {\bf C2} (1998) 247.
\bibitem{omegaphoto} ZEUS Collab., M.~Derrick et al., Z.Phys. {\bf C73} (1996) 1, 73.
\bibitem{phiphoto} ZEUS Collab., M.~Derrick et al., Phys.Lett. {\bf B377} (1996) 259.
\bibitem{jpsiphoto} H1 Collab., C.~Adloff et al., Phys.Lett. {\bf B483} (2000) 23;  Paper 437, ICHEP200, Osaka, July 2000.
\bibitem{psiphoto} Abstracts 987, 985, ICHEP200, Osaka, July 2000.
\bibitem{rhoelec} H1 Collab., C.~Adloff et al., Eur.Phys.J. {\bf C13} (2000) 371;  Paper 439, ICHEP200, Osaka, July 2000.
\bibitem{omegaelec} ZEUS Collab., J.~Breitweg et al., accepted by Phys.Lett. B-PLB 16283.
\bibitem{phielec} H1 Collab., C.~Adloff et al., Phys.Lett. {\bf B483} (2000) 360; Abstract 793 at ICHEP98, Vancouver, July 1998.
\bibitem{jpsielec} H1 Collab., C.~Adloff et al., Eur.Phys.J. {\bf C10} (1999) 373; Paper 438, ICHEP200, Osaka, July 2000.
\bibitem{hight} Paper 442, ICHEP200, Osaka, July 2000.
\bibitem{talk} B.~Mellado, talk 02c-01 at ICHEP2000, Osaka, July 2000.
\bibitem{clerbaux} B.~Clerbaux, hep-ph/9908519.
\bibitem{naroska} B.~Naroska, VM Production at HERA, XXXVth Rencontres de Moriond, March 2000, Les Arcs, France.


\end{thebibliography}
\end{document}